\documentclass{PoS}

\def\mpi2{m_\pi^2}
\def\mK2{m_K^2}

\newcommand{\bea}{\begin{eqnarray}}
\newcommand{\eea}{\end{eqnarray}}
\newcommand{\be}{\begin{equation}}
\newcommand{\ee}{\end{equation}}

\def\lsim{\raise0.3ex\hbox{$<$\kern-0.75em\raise-1.1ex\hbox{$\sim$}}}
\newsavebox{\DERIVBOXZLM}
\savebox{\DERIVBOXZLM}[2.5em]{$\Longrightarrow\hspace{-1.5em}
\raisebox{.2ex}{*}
\hspace{-.7em}\raisebox{-.8ex}{\scriptsize lm}\hspace{.7em}$}

\usepackage{epsf}

\def\chpt{\raise0.4ex\hbox{$\chi$}PT}
\def\schpt{S\raise0.4ex\hbox{$\chi$}PT}
\def\rschpt{rS\raise0.4ex\hbox{$\chi$}PT}
\def\figref#1{Fig.~\ref{fig:#1}}
\def\Figref#1{Figure~\ref{fig:#1}}

\def\gtwid{{\,\raise.3ex\hbox{$>$\kern-.75em\lower1ex\hbox{$\sim$}}\,}}
\def\ltwid{{\,\raise.3ex\hbox{$<$\kern-.75em\lower1ex\hbox{$\sim$}}\,}}

\def\ie{{\it i.e.},\ }
\def\eg{{\it e.g.},\ }
\def\et{{\it et al.}}

\def\vs{{\it vs.}\ }

\def\cM{{\cal M}}

\def\rcite#1{Ref.~\cite{#1}}

\def\eqn#1{\label{eq:#1}}

\def\eq#1{Eq.~(\ref{eq:#1})}

\title{Electromagnetic contributions to pseudoscalar masses}

\ShortTitle{Electromagnetic contributions to pseudoscalar masses}

\author{S.~Basak$^a$, A.~Bazavov$^b$, \speaker{C.~Bernard}$^c$, C.~DeTar$^d$, E.~Freeland$^e$, W.~Freeman$^f$,
 J.~Foley$^d$, Steven~Gottlieb$^g$, U.M.~Heller$^h$, J.E.~Hetrick$^i$, J.~Laiho$^{j,k}$, L.~Levkova$^{l,d}$,
 M.~Oktay$^d$, J.~Osborn$^m$, R.L.~Sugar$^n$, A.~Torok$^g$, D.~Toussaint$^l$, R.S.~Van~de~Water$^{o,b}$, R.~Zhou$^g$ \,(MILC Collaboration)\\

$^a$ NISER, Bhubaneswar, Orissa 751005, India      \\
$^b$ Physics Department, Brookhaven National Laboratory, Upton, NY 11973, USA\\
$^c$ Department of Physics, Washington University, St. Louis, MO 63130, USA\\
$^d$ Department of Physics and Astronomy, University of Utah, Salt Lake City, UT 84112, USA\\
$^e$ Department of Physics, Benedictine University, Lisle, IL 60532, USA\\
$^f$ Department of Physics, George Washington University, Washington, DC 20037, USA\\
$^g$ Department of Physics, Indiana University, Bloomington, IN 47405, USA\\
$^h$ American Physical Society, One Research Road, Ridge, NY 11961, USA\\
$^i$ Physics Department, University of the Pacific, Stockton, CA 95211, USA\\
$^j$ SUPA, School of Physics and Astronomy, University of Glasgow, Glasgow, UK\\
$^k$ Department of Physics, Syracuse University, Syracuse, NY  13244, USA\\
$^l$ Physics Department, University of Arizona Tucson, AZ 85721, USA\\
$^m$ ALCF, Argonne National Laboratory, Argonne, IL 60439, USA\\
$^n$ Physics Department, University of California, Santa Barbara, CA 93106, USA\\
$^o$ Theoretical Physics Department, Fermi National Accelerator Laboratory, Batavia 60510, USA\\
E-mail: \email{cb@wustl.edu}
}
\abstract{We report on the calculation  by the MILC Collaboration of 
the electromagnetic effects on kaon and pion masses.  These
masses are computed in QCD with dynamical (asqtad staggered) quarks plus quenched photons at 
three lattice spacings varying from 0.12 to 0.06 fm.
The masses are fit to staggered chiral perturbation theory 
with NLO electromagnetic terms, as well as analytic terms at higher order.
We extrapolate the results to physical light-quark masses
and to the continuum limit. At the current stage of the analysis,
most, but not all, of the systematic errors have been estimated. 
The main goal is the comparison of kaon electromagnetic splittings to 
those of the pion, \ie an
evaluation of the corrections to ``Dashen's theorem.''  This in turn will 
allow us to
significantly reduce the systematic errors in our determination of $m_u/m_d$.}

\FullConference{The 7th International Workshop on Chiral Dynamics,\\
                August 6 -10, 2012\\
                Jefferson Lab, Newport News, Virginia, USA}

\begin{document}

{\it Introduction. --}
The disentangling of electromagnetic (EM) and isospin-violating
effects in the kaon and pion systems is a long-standing problem.
Understanding these effects is crucial for computing light quark masses,
which are fundamental parameters in the Standard Model and  important for
phenomenology.
Indeed, the size of the EM contributions to the kaon masses
is the largest uncertainty in 
determinations of $m_u/m_d$ from the lattice \cite{Colangelo:2010et}, and in particular
in our calculations \cite{qrat}.
The contributions have until recently been taken from a variety of
phenomenological
estimates, and therefore have quite large and not well controlled errors.
We have been working on reducing these uncertainties for some
time by calculating the EM effects directly on the lattice; 
progress has been reported previously in Refs.~\cite{basak,levkova}.

The error in $m_u/m_d$ is dominated by the error in the mass difference
$(M^2_{K^\pm}-M^2_{K^0})^\gamma$, where 
$\gamma$ denotes the total EM contribution, \ie the difference between
the value of a quantity in the presence of electromagnetism, and 
its value in a world in which all EM charges, 
both of valence and of sea quarks, are turned off.
One may try to relate 
this difference to the much better understood difference
in the pion system, $(M^2_{\pi^\pm}-M^2_{\pi^0})^\gamma$.
To lowest order (LO) in chiral perturbation theory (\chpt) these EM splittings
are in fact the same; this observation is known as Dashen's theorem \cite{dashen}.
We aim to calculate on the lattice the corrections to Dashen's theorem, which
may be parameterized by
\vspace{-1.5mm}
\begin{equation}
\eqn{eps-def}
 (M^2_{K^\pm}-M^2_{K^0})^\gamma =(1+\epsilon) (M^2_{\pi^\pm}-M^2_{\pi^0})^\gamma\ .
\vspace{-2mm}
\end{equation}
Our computations employ full QCD but {\it quenched}\/ photons.  As
pointed out in \rcite{BD}, however, the EM-quenching effects on $\epsilon$ may be calculated 
and corrected to NLO in \chpt, with controlled errors.   
Similarly, squared mass differences of the form
$(M^2_{P}-M^2_{P'})^\gamma$ are calculable with controlled errors
in our setup, where $P$ is any light pseudoscalar meson 
and $P'$ is the corresponding 
meson made from neutral valence quarks with the same masses as those in $P$. 
We also compute the EM effects on the $K^0$ alone,
namely $(M^2_{K^0})^\gamma$.  
In this case, however, the quenching effects are not calculable in \chpt, and uncontrolled
errors remain.  For  $m_u/m_d$, the uncertainty coming from $(M^2_{K^0})^\gamma$ is, fortunately,
subdominant.

In the pion system, isospin-violating effects on the mass splitting are known 
to be small (see, \eg \rcite{Colangelo:2010et}),
so the experimental splitting is almost completely electromagnetic:
$ (M^2_{\pi^\pm}-M^2_{\pi^0})^{\rm expt} \approx (M^2_{\pi^\pm}-M^2_{\pi^0})^\gamma$.
It would be costly to simulate the true $\pi^0$, which has EM disconnected
diagrams even in the isospin limit.  Instead, we simulate a ``$\pi_0$'' whose squared mass
is a simple average of the squared masses of $u\bar u$ and $d\bar d$ mesons, computed  
with connected diagrams only.  Because all EM contributions to neutral mesons vanish in 
the chiral limit,
both the true $(M^2_{\pi^0})^\gamma$ and our  $(M^2_{``\pi^0"})^\gamma$ are 
small in any case.
From Zweig-rule considerations, we suspect that 
the disconnected contribution is smaller still.
We estimate
\vspace{-2.5mm}
\begin{equation}
\eqn{pi0-EM}
 (M^2_{\pi^0})^\gamma \sim (M^2_{``\pi^0"}-M^2_{\pi'})^\gamma\ .
\vspace{-2mm}
\end{equation}
Since there are errors coming both from quenched electromagnetism and from
the neglect of disconnected diagrams, one might expect $\sim\!100\%$ errors 
in this estimate, however.
On the other hand, our calculation of the EM pion splitting, using
$ (M^2_{\pi^\pm}-M^2_{\pi^0})^\gamma \approx (M^2_{\pi^\pm}-M^2_{``\pi^0"})^\gamma$
suffers only from the neglect of disconnected diagrams, not from 
uncontrolled quenching effects. As a rough estimate of the former
we take 50\% of $(M^2_{\pi^0})^\gamma $, calculated through \eq{pi0-EM}.

{\it Chiral Perturbation Theory. --}
We fit our lattice data to expressions from (partially quenched) 
staggered chiral perturbation theory (\schpt) in order
to extrapolate to the physical light quark masses and to the continuum.  
We consider Goldstone (taste $\xi_5$) pseudoscalar mesons composed of valence quark $x$ and valence antiquark $y$,
with masses $m_x$ and $m_y$.
Let $M_{xy,5}$ be the mass of such a meson with valence-quark charges
$q_x$ and $q_y$ ($q_{xy}\equiv q_x-q_y$ is the meson charge), and let $\Delta M^2_{xy,5}$ be the
squared-mass splitting $\Delta M^2_{xy,5}\equiv M^2_{xy,5}- M^2_{x'y',5}$, 
where  the primes in the second subscript
indicate that the valence-quark charges are set to zero. 
To NLO in \schpt, $\Delta M^2_{xy,5}$ is given by \cite{schpt}:
\vspace{-4mm}
\begin{eqnarray}
\Delta M^2_{xy,5}&=&q^2_{xy}{\delta_{EM}}-\frac{1}{16\pi^2}e^2q^2_{xy}\cM^2_{xy,5}\left
[3\ln(\cM^2_{xy,5}/\Lambda^2_\chi)-4\right]\nonumber \\
&&-\frac{2{\delta_{EM}}}{16\pi^2f^2}\frac{1}{16}\sum_{\sigma,\xi}\left[q_{x\sigma}q_{xy}\cM^2_{x\sigma,\xi}\ln(\cM^2_{x\sigma,\xi})
-q_{y\sigma}q_{xy}\cM^2_{y\sigma,\xi}\ln(\cM^2_{y\sigma,\xi})\right]
\label{eq:fit}\\
&& \hspace{-21mm}+c_1 q_{xy}^2a^2+c_2q_{xy}^2(2m_l+m_s)+c_3(q_x^2+q_y^2)(m_x+m_y)+c_4q_{xy}^2(m_x+m_y)+c_5(q_x^2m_x+q_y^2m_y) \nonumber
\vspace{-6mm}
\end{eqnarray}
where $\delta_{EM}$ is a low-energy 
constant (LEC), $\sigma$ runs over the sea quarks,
$\xi$ runs over the staggered tastes, $c_i$ 
are the LECs at NLO, and $m_l$, $m_s$ are the light and strange sea-quark masses.
At this order, the meson masses 
denoted  by $\cM$ on the right hand side may be
taken to be the tree-level masses in the absence of electromagnetism.
Note that \rcite{schpt} mentions an additional NLO analytic term proportional to $(q_x^2+q_y^2)a^2$; this
is not possible for a Goldstone splitting because it does not vanish 
as $m_x+m_y$ for small quark masses in the limit $q_y=q_x$, where EM effects are
chirally symmetric.

Our statistical errors in $\Delta M^2_{xy,5}$ are
$\sim\!0.3\%$ for charged mesons and $\sim\!1.0\%$ for neutral mesons.  It is
clear that NLO \schpt\ cannot be expected to give a good description of
the splittings at that level of precision.  For reasonable fits, NNLO terms are 
needed. The \schpt\ logarithms have not
been calculated at that order, but we add all possible analytic terms.
This may be justified by noting that the NNLO logarithms will be small
at low mass, where the extrapolation is performed, and will be well approximated
by analytic terms in the region near $m_s$.

To allow for finite-volume (FV) effects, we include standard terms
dependent on $m_\pi L_s$ from EM tadpoles ($L_s$ is the spatial lattice size), as well as 
an empirical EM finite-volume correction of the form $f_vq_{xy}^2/L_s^2$ used previously in Ref.~\cite{BMW},  
where $f_v$ is a constant.
However, as discussed below, our measured finite-volume effects are rather small at present, and including or omitting the finite-volume terms from the fits makes little difference in the final results.

{\it Lattice setup. --}
We calculate the meson spectrum in quenched EM backgrounds
on a set of asqtad ensembles with 2+1 flavors and $0.12\;{\rm fm}\gtwid a\gtwid 0.06\;{\rm fm}$.  
See \rcite{levkova} for a table of lattice parameters.
The valence quarks have charges $\pm2/3 e$, $\pm1/3 e$ or 0,
where $e= e_{\rm phys}$, $2e_{\rm phys}$, or $3e_{\rm phys}$ ($e_{\rm phys}$
is the physical electron charge).
For the results reported here,
only the $e=e_{\rm phys}$ data is used.

\begin{figure}[b]
\epsfxsize=0.55\textwidth
\begin{center}
\epsfbox{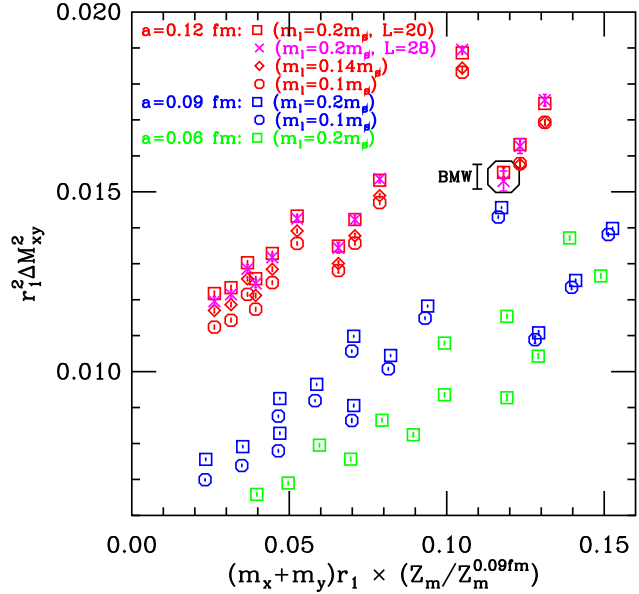}
\end{center}
\vspace{-6mm}
\caption{A sampling of our partially quenched data in $r_1$ units 
for EM splittings of  pseudoscalar mesons with charge 
$\pm e_{\rm phys}$, plotted versus the sum of the valence-quark masses.
For clarity, only about a quarter of the data is shown.
The red squares and magenta crosses show results for the two ensembles 
that differ only by the spatial volume: $20^3$ and $28^3$, respectively.
The vertical black bar 
labeled ``BMW'' shows the expected difference for kaons 
between these two volumes,
based on the results from the BMW collaboration \cite{BMW}.  Next to it,
the two points encircled in black are our ``kaon-like'' points for the volumes.
}
\label{fig:FV}
\vspace{-3mm}
\end{figure}

\Figref{FV} shows some of our partially quenched data for the splittings of
physically charged $\pi^\pm$ and $K^\pm$ mesons.
We investigate the FV effects on
two ensembles with $a\approx 0.12$ fm, 
$am_l=0.01$, and $am_s=0.05$, but different volumes ($L\equiv L_s/a= 20$ and $28$).  
The figure shows a comparison of the FV effect seen in our data 
with the difference between the two volumes expected from the results 
of the BMW Collaboration \cite{BMW}.  The effect we see is smaller:
0.35(45) of the difference expected from BMW.  
However, our larger-volume ensemble has only
274 configurations, leading to the rather large errors in the comparison.
We are currently increasing the 
statistics on that ensemble 
in hopes of clarifying the issue.

We have measured the taste splitting between the Goldstone 
pion and the other local pion on our ensembles. The amount of taste violation
caused by high-momentum photons is nonnegligible for mesons
made from quarks with higher-than-physical charges, although it is always 
significantly smaller than that caused by high-momentum gluons.
The fit function Eq.~(\ref{eq:fit}) is based
on the neglect of taste violations caused by photons; that is the reason
that we  focus here only on the data with physical quark charges.  
Given that photon-induced taste 
violations are relatively small, however, one could 
expand the fit function in powers of $\alpha_{EM}=e^2/(4\pi)$. Thus, 
inclusion of $\alpha^2_{EM}$ analytic terms to the fit function should allow the higher-charge data to
be fit. That approach seems to work, and will be explored more in the future.
For more details on EM taste-violating effects, see \rcite{levkova}.

\begin{figure}[t]
\epsfxsize=0.6\textwidth
\begin{center}
\epsfbox{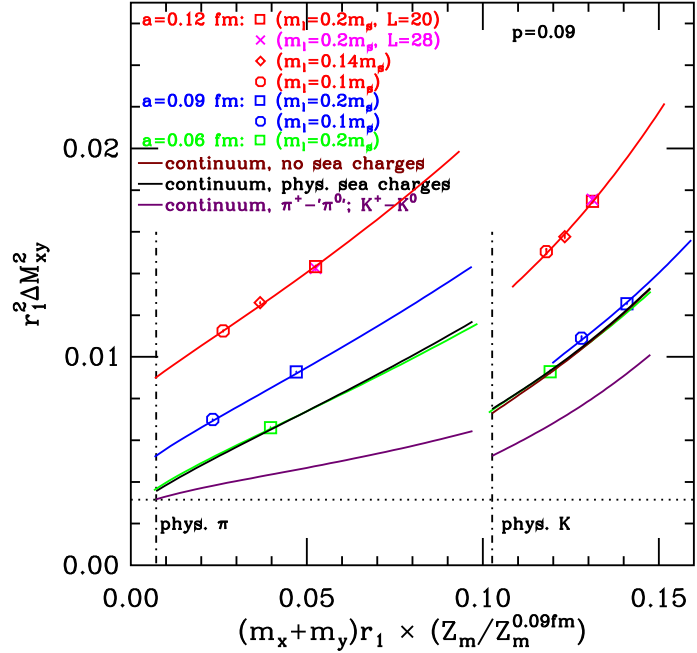}
\end{center}
\vspace{-6mm}
\caption{Typical \schpt\ fit to  the squared-mass EM splitting $\Delta M^2$ \vs\ the
sum of the valence-quark masses. 
Only a small subset of the charged-meson data
is shown.
The red, blue and green curves correspond to three different lattice spacings.
The brown and black curves are continuum limits for $\Delta M^2$, without or with the correction
from \chpt\ for physical sea-quark charges.  The purple
curves are the continuum limits for the $K^+$--$K^0$ splitting (right),
and the $\pi^+$--$``\pi^{0}"$ splitting (left).
\label{fig:EMfit}
\vspace{-2mm}
}
\end{figure}

{\it Results and Outlook. --}
\Figref{EMfit} shows a typical fit of  our data for $\Delta M^2$
with physical quark charges to \eq{fit} (with added analytic NNLO terms).
We fit partially quenched charged- and neutral-meson data simultaneously, 
but only the (unitary or approximately
unitary) charged-meson data is shown in the plot. 
This fit has 55 data points and 26 parameters; other fits have as many as 120 data points, and from
20 to 30 fit parameters, depending
on how many of the NNLO terms are included, and whether small variations with $a^2$
of the LO and NLO low-energy constants are allowed. 
The covariance matrix of the data is nearly singular, and the
statistics are insufficient to determine it with enough precision to
yield good correlated fits, so almost all fits currently used are uncorrelated. 
The fit shown has an (uncorrelated) $p$ value of 0.09.
We note that what appear to be big discretization effects 
are actually due in large part to 
mistunings of the strange-quark mass, which is off by about 50\% on the
$a=0.12$ fm ensembles and 25\% on the $a=0.09$ fm ensembles, but only by 2\% on the $0.06$ fm ensemble.

The black and brown lines in \figref{EMfit} show the fit after setting valence and sea masses equal,
adjusting $m_s$ to its physical value, and extrapolating to the continuum.  
The black lines adjust the sea charges to
their physical values using NLO \chpt,
while the brown line keeps the sea quarks uncharged.
In the pion case, 
the adjustment vanishes identically, so no brown line is visible.  In the 
kaon case, the adjustment is a very small correction.  From the black lines for the $\pi^+$ and $K^+$,
we subtract the corresponding results for the neutral mesons, $``\pi^0"$ and $K^0$, giving
the purple lines. Results for $(M_{\pi^+}^2 - M_{``\pi^0"}^2)^\gamma$ and
$(M_{K^+}^2 - M_{K^0}^2)^\gamma$ are then obtained from the intersections of the purple lines and
the vertical dashed-dotted lines that give the location of the physical point for each meson.   
The excellent agreement of the result for $(M_{\pi^+}^2 - M_{``\pi^0"}^2)^\gamma$ and the experimental
pion splitting (horizontal dotted line) is accidental, since our result has roughly 20\% total error.

We find the following preliminary results:
\vspace{-0.5mm}
\begin{eqnarray}
\eqn{results}
(M_{K^+}^2 - M_{K^0}^2)^\gamma &=& 2100(90)(250)\hspace{1mm} {\rm MeV}^2\;, \hspace{16mm}
(M_{K^0}^2)^\gamma = 901(8)(9)(?)\hspace{1mm} {\rm MeV}^2\nonumber \\
(M_{\pi^+}^2 - M_{``\pi^0"}^2)^\gamma &=& 1270(90)(230)(80)\hspace{1mm} {\rm MeV}^2\;,\hspace{7mm}
(M_{``\pi^0"}^2)^\gamma = 157.8(1.4)(1.7)(?)\hspace{1mm} {\rm MeV}^2 \nonumber\\
\epsilon&=&0.65(7)(14)(10)
\vspace{-5mm}
\end{eqnarray}
The first two errors in each case are statistical and lattice systematic uncertainties. 
 The latter error comes largely from the effects of changing the assumptions entering
into the chiral/continuum fit.  Note, however,
that finite-volume errors are {\it not}\/ included at present. We expect that ultimately they will
be a significant, but subdominant, source of error.
The ``$?$'' for $(M_{K^0}^2)^\gamma$ and $(M_{``\pi^0"}^2)^\gamma$ represent the effect
of EM quenching
and, for  $(M_{``\pi^0"}^2)^\gamma$, the effect of neglected disconnected diagrams. These errors are
likely to be much larger than the small quoted errors. For $(M_{\pi^+}^2 - M_{``\pi^0"}^2)^\gamma$ and
$\epsilon$ the third
error is a rough guess of the effect of neglecting disconnected diagrams, which we estimate
by 50\% of the result for $(M_{``\pi^0"}^2)^\gamma$. If we redefine $\epsilon$ by replacing
our  computation of the pion EM splitting in \eq{eps-def} with the experimental
splitting, we get $\epsilon=0.66(7)(20)$,
which has larger chiral/continuum-extrapolation errors, but no error from  neglecting disconnected contributions.
Our result for $\epsilon$ is compatible with results from other groups
\cite{Blum:2010ym,BMW}.
Using our values for $\epsilon$ and $(M_{K^0}^2)^\gamma$ (assuming 100\% EM-quenching
error in the latter quantity), 
our preliminary estimate for the EM uncertainty in $m_u/m_d$ is reduced by approximately
a factor of two \cite{doug} from our previous error \cite{qrat}.

We are currently finishing the analysis of two more-chiral ensembles at $a\approx 0.06$ fm in order
to improve the chiral and continuum extrapolations, and are increasing the 
statistics for our finite-volume study. A ``second-generation'' calculation on HISQ ensembles,
including ones at physical quark masses, is now
beginning, and promises significant reductions in systematic errors.  Calculations with dynamical
EM effects included are also being contemplated.

{\bf Acknowledgments:} We thank Laurent Lellouch and Taku Izubuchi for useful discussions.  Computations for this work were carried out with resources provided by the USQCD
Collaboration, the Argonne Leadership Computing Facility, and the National Energy Research Scientific 
Computing Center, which are funded by the Office of Science of the U.S.
Department of Energy; and with resources provided by the National Center for Atmospheric
Research, the National Institute for Computational Science, the Pittsburgh Supercomputer
Center, the San Diego Supercomputer Center, and the Texas Advanced Computer Center,
which are funded through the National Science Foundation's XSEDE Program. This work
was supported in part by the U.S. Department of Energy under Grants DE-FG02-91ER-
40628, DE-FG02-91ER-40661, DE-FG02-04ER-41298, and DE-FC02-06ER41446; and by
the National Science Foundation under Grants PHY07-57333, PHY07-03296, PHY07-57035,
PHY07-04171, PHY09-03571, PHY09-70137, and PHY10-67881. This manuscript has been
co-authored by an employee of Brookhaven Science Associates, LLC, under Contract No.
DE-AC02-98CH10886 with the U.S. Department of Energy. Fermilab is operated by Fermi
Research Alliance, LLC, under Contract No. DE-AC02-07CH11359 with the U.S. Department of Energy. 
For this work we employ QUDA \cite{quda}.


\vspace{-4mm}
\begin{thebibliography}{99}

\vspace{-3mm}

\bibitem{Colangelo:2010et} 
  G.~Colangelo, 
{\it et al.},
  Eur.\ Phys.\ J.\ C {\bf 71}, 1695 (2011)
  [arXiv:1011.4408].


\bibitem{qrat} 
A. Bazavov {\it et al.}, 
Rev. Mod. Phys. 82, 1349 (2010) [arXiv:0903.3598]; 
A. Bazavov {\it et al}.
[MILC], PoS LATTICE {\bf 2009}, 079 (2009) [arXiv:0910.3618];
C. Aubin \et,
[MILC], Phys. Rev. D 70, 114501 (2004) [hep-lat/0407028]. 

\bibitem{basak} S. Basak {\it et al.} [MILC], PoS LATTICE {\bf 2008}, 127 (2008) [arXiv:0812.4486];
A.~Torok {\it et al.},
  PoS LATTICE {\bf 2010}, 127 (2010).

\bibitem{levkova}
  S.~Basak {\it et al.}  [MILC],
  PoS LATTICE {\bf 2012}, 137 (2012)
  [arXiv:1210.8157].


\bibitem{dashen} R. Dashen, Phys. Rev. {\bf 183}, 1245 (1969). 

\bibitem{BD}
J.~Bijnens and N.~Danielsson,
  Phys.\ Rev.\ D {\bf 75}, 014505 (2007)
  [hep-lat/0610127].

\bibitem{schpt} C.~Bernard and E.~D.~Freeland,
  PoS LATTICE {\bf 2010}, 084 (2010)
  [arXiv:1011.3994].

\bibitem{BMW}
  A.~Portelli {\it et al.}, [BMW],
  PoS LATTICE {\bf 2011}, 136 (2011)
  [arXiv:1201.2787] and
  PoS LATTICE {\bf 2010}, 121 (2010)
  [arXiv:1011.4189].


\bibitem{Blum:2010ym} 
  T.~Blum {\it et al.},
  Phys.\ Rev.\ D {\bf 82}, 094508 (2010)
  [arXiv:1006.1311].


\bibitem{doug} 
  A.~Bazavov {\it et al.}  [Fermilab Lattice and MILC],
  PoS LATTICE {\bf 2012}, 159 (2012)
  [arXiv:1210.8431].
%
\bibitem{quda} M.A.\ Clark {\it et al.}, Comput.\ Phys.\ Commun.\ 181, 1517 (2010) [arXiv:0911.3191]; R.\ Babich {\it et al.}, 
[arXiv:1109.2935].
\end{thebibliography}
\end{document}